\documentclass[showpacs]{revtex4}
\usepackage{graphicx}
\usepackage{dcolumn}
\usepackage{bm}

\begin{document}

\title{Differences between normal and shuffled texts: structural properties of weighted networks}
\author{A. P. Masucci}
\author{G. J. Rodgers}%
\affiliation{%
Department of Mathematical Sciences, Brunel University,
Uxbridge, Middlesex, UB8 3PH, United Kingdom}%

\date{\today}
\begin{abstract}
In this paper we deal with the structural properties of weighted networks.  Starting from an empirical analysis of a linguistic network, we analyse the differences between the statistical properties of a real and a shuffled network and we show that the scale free degree distribution and the scale free weight distribution are induced by the scale free strength distribution, that is  Zipf's law. We test the result on  a scientific collaboration network, that is a social network, and we define a measure, the vertex selectivity, that can easily distinguish a real network from a shuffled network. We prove, via an ad-hoc stochastic growing network with second order correlations, that this measure can effectively capture the correlations within the topology of the network.
 \end{abstract}
 \maketitle

\section{\label{sec:level1}Introduction.\protect}

Following the seminal work of Barabasi \emph{et al}. \cite{1}, the scientific community has put a lot of effort into the study of network theory. A network is a  collection of vertices connected by edges. Often vertices represent natural elements or events and the edges relations by which those elements or events are connected.  As it is a simple framework, network theory can be applied to a wide variety of natural phenomena, including social sciences \cite{2},  biology, genetics \cite{3},  geology \cite{18} and linguistics \cite{4,10}. The extraordinary similarities that such different phenomena display when represented by network theory  provide us with  the opportunity of finding common principles for the organisation of many elements in nature.

 We begin this work with the empirical study of a linguistic network built from the novel of Herman Melville, $Moby$ $Dick$ \cite{5}. This is a multi-directed Eulerian network \cite{6}, based on the relation of adjacency, where the tokens of the novel, words and punctuation, are the vertices, and two vertices are linked if the tokens they represent are adjacent in the text. We then analyse a scientific collaboration network, that is a network in which vertices are the authors of scientific published papers related to network theory \cite{7}, and two vertices are connected if the authors they represent coauthored the same paper.

 These networks are suitable for analysis as weighted network \cite{8}, that is they are networks in which pairs of vertices represent events that are related more than once in the  phenomenon. This multiple relation can be expressed by a weight on their mutual link, the weight representing the number of times the relation is repeated.  To completely describe the network, we introduce the weighted adjacency matrix $W=\{w_{ij}\}$, that is a matrix whose elements $w_{ij}$ represent the number of  links connecting vertex $i$ to vertex $j$. In the case of an undirected network, such as the scientific collaboration network, this matrix is symmetric, and there is no difference between out and in vertices properties. In the case of a directed network, such as the  language network, the matrix is not symmetric, and the out and in vertices properties are generally different. We define the out-degree and in-degree $k_i^{out/in}$ of vertex $i$ as the number of its out and in nearest neighbours
and we have $k_i^{out/in}\equiv\sum_j\Theta(w_{ij/ji}-\frac{1}{2})$, where $\Theta(x)$ is the Heaviside function. We
define the out-strength and the in-strength $s_i^{out/in}$ of the vertex $i$ as the number of its outgoing and incoming links, that is $s_i^{out/in}\equiv\sum_jw_{ij/ji}$.

To understand the properties of complex system, it is common to consider random systems as a null hypothesis. We  apply this concept to our networks by considering the classical measures (weight, strength, degree, clustering coefficient, nearest neighbours degree) applied to the real networks, and then again after  shuffling the vertices of the network. In the case of the linguistic network the strength of a vertex is equivalent to the frequency of the token the vertex represents. The shuffling operation, in this case, consists of shuffling the tokens of the novel. This operation doesn't alter the frequency of the tokens and so the strength distribution remains unchanged. In particular, if we call $f(s)$, the frequency of a token with strength $s$ and $r$ the rank of a word in the definition given by Zipf \cite{9}, we have that $r=\sum_{s=1}^{s_{Max}}f(s)$, that is a direct relation between the rank and the strength of a token. Hence, as is already well known, the shuffling operation doesn't alter  Zipf's law. In the same way, in the case of the scientific collaboration network, the strength of a vertex represents the number of papers an author published. The shuffling operation preserves this number and hence the strength distribution remains unchanged.

In this paper we will show that, in both our networks, the degree distribution is  not altered by the shuffling process, and seems to be determined by the strength distribution. We then introduce a new measure, the \emph{vertex selectivity}, that reflects the correlations of the networks and is greatly changed in the shuffling process. We test this measure with an ad-hoc stochastic network characterised by short range correlations.

\section{Analysis of Moby Dick}

\textit{Moby Dick} is a text that is large enough to be suitable for a statistical
analysis and, since it is  considered by the critics and by the people a very well written text, we are confident that the statistical empirical laws arising from it are  characteristic of laws of language \cite{11}. After shuffling the tokens of the novel, we will compare some empirical measures between the shuffled and the original text.

$Moby$ $Dick$  has a vocabulary $V$ of 17169 tokens, with a total size $N$ of 264978 tokens. Since the network is Eulerian \cite{6} we have that $s^{out}_i=s^{in}_i=\frac{s_i}{2}$, for every vertex $i$. The  average strength is $<s>\approx30.87$, the average out-degree is $<k_{out}>\approx 6.54$.

\begin{figure}[!htbp]\center
         \includegraphics[width=0.49\textwidth]{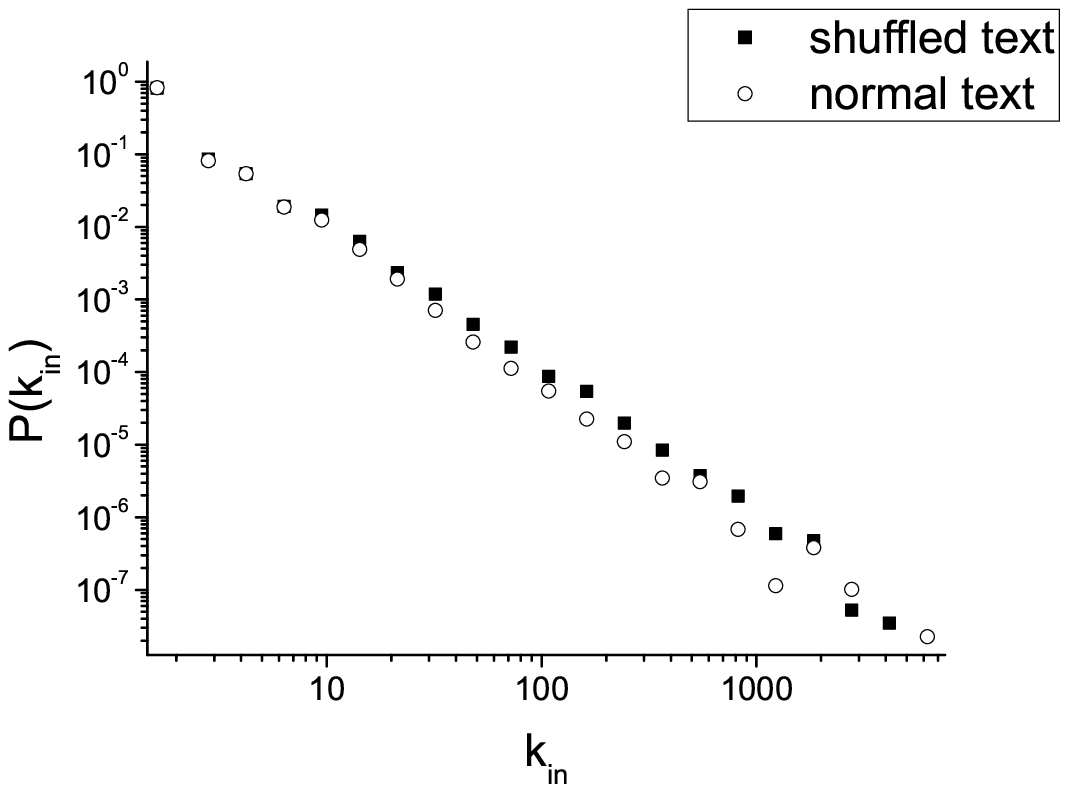}
         \includegraphics[width=0.49\textwidth]{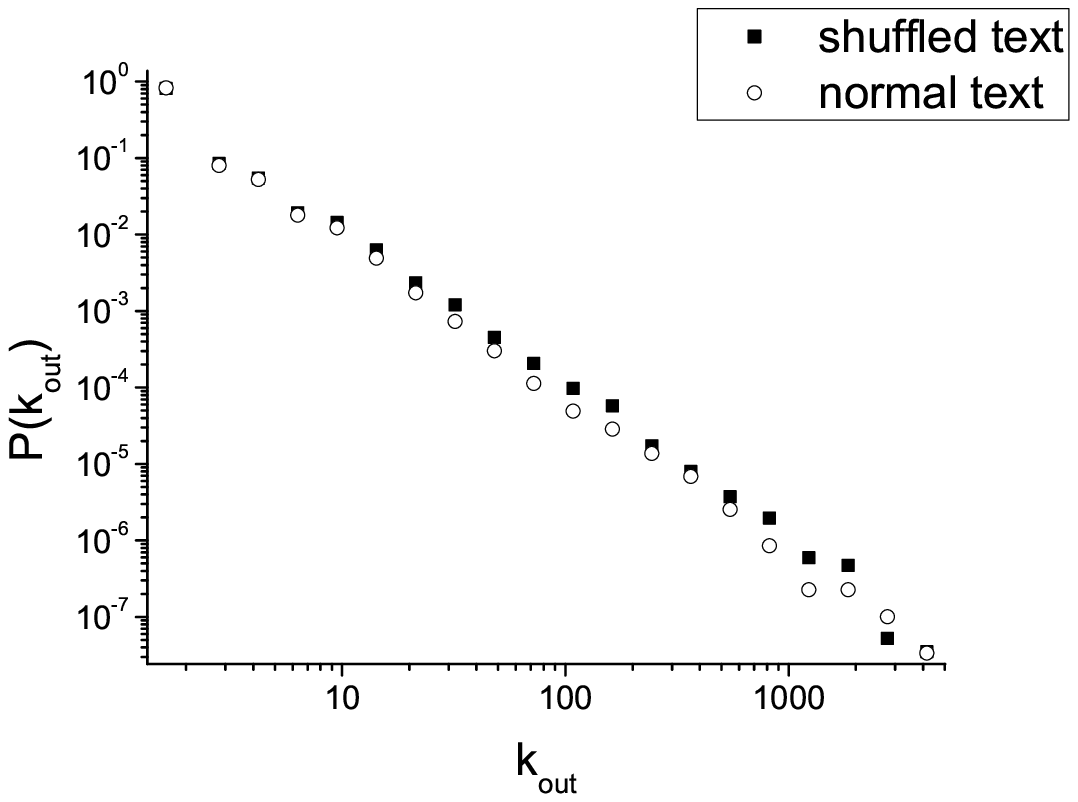}
          \includegraphics[width=0.49\textwidth]{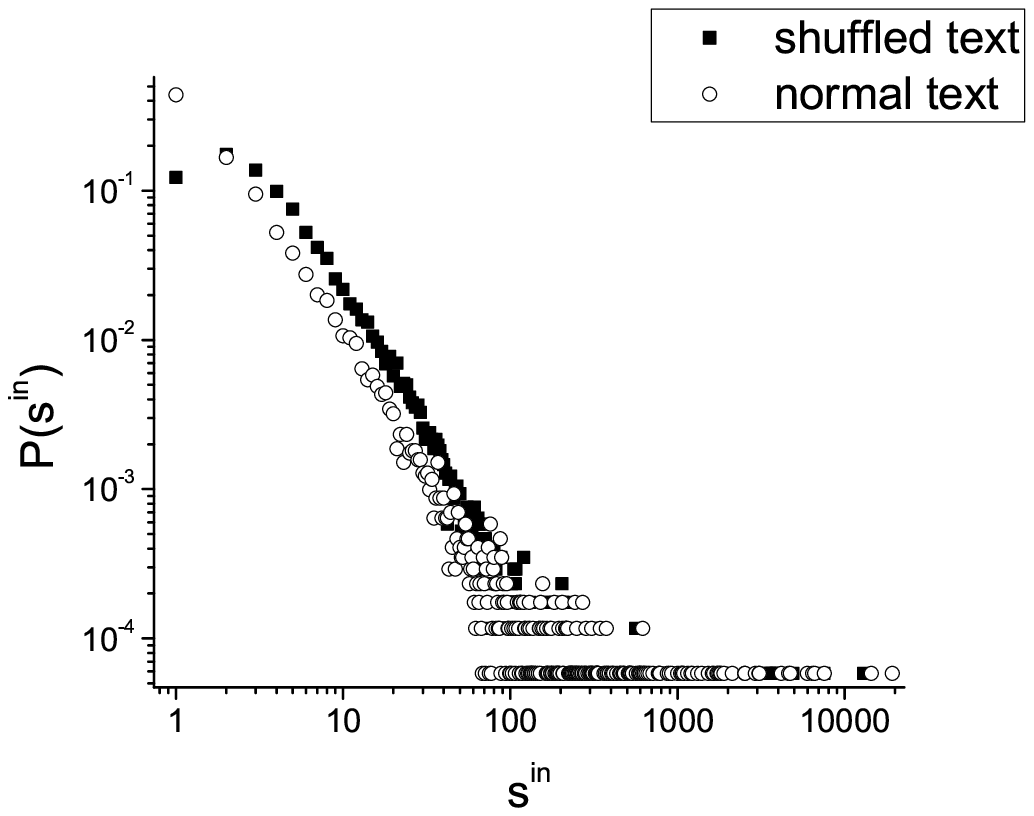}
         \includegraphics[width=0.49\textwidth]{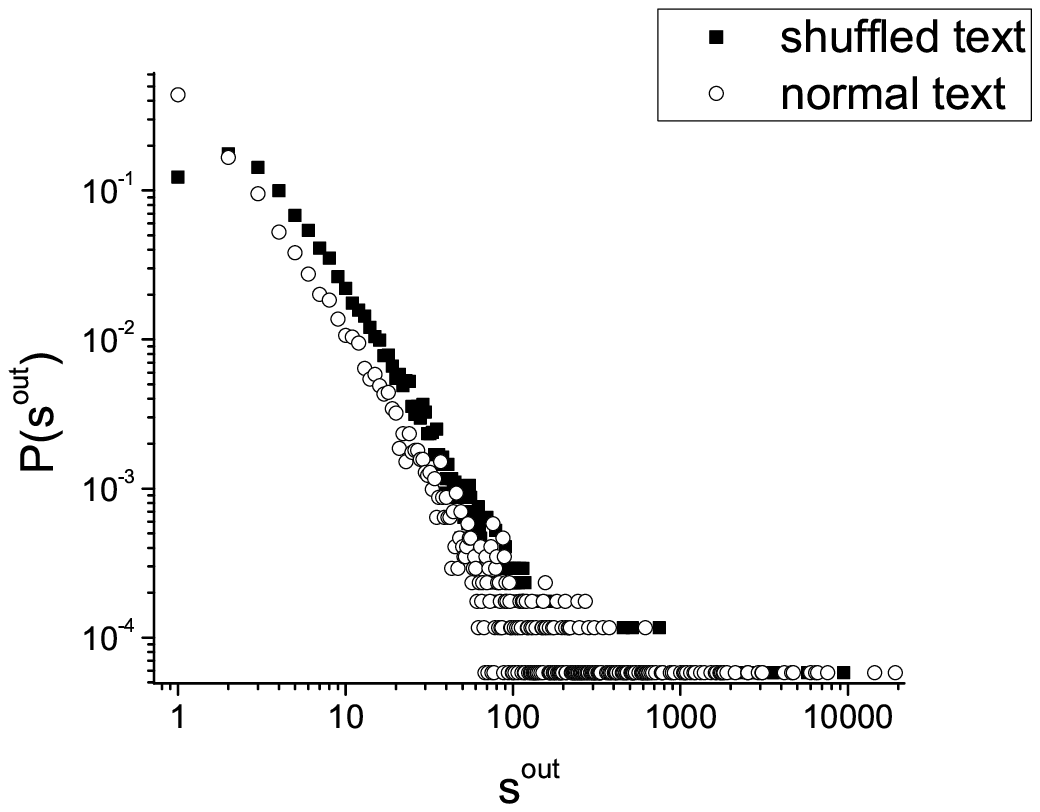}

\caption{\label{f1} Top panels: the out and in-degree distributions of \textit{Moby Dick} compared to the distributions obtained after shuffling the tokens of the novel, preserving the strength of the vertices. The result implies that the scale free degree distribution is determined by the scale free strength distribution. Bottom panels: the out and in-strength distributions of \textit{Moby Dick} compared to the distributions obtained after shuffling the links between the tokens of the novel, preserving the degree of the vertices. The result implies that the scale free strength distribution is not determined by the scale free degree distribution.}
\end{figure}

\begin{figure}[!htbp]\center
         \includegraphics[width=0.58\textwidth]{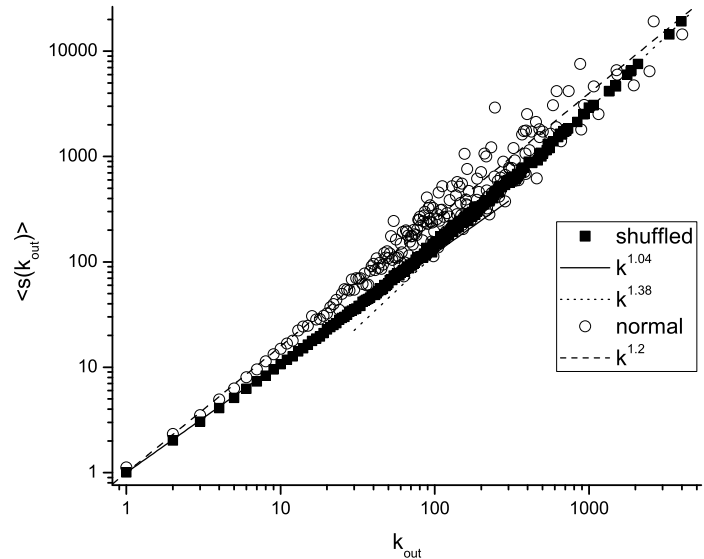}

\caption{\label{f0} Strength of the vertices averaged on the out-degree versus the out-degree for \textit{Moby Dick} compared to the same measure obtained after shuffling the tokens of the novel, preserving the strength of the vertices.}
\end{figure}

\begin{figure}[!htbp]\center
         \includegraphics[width=0.49\textwidth]{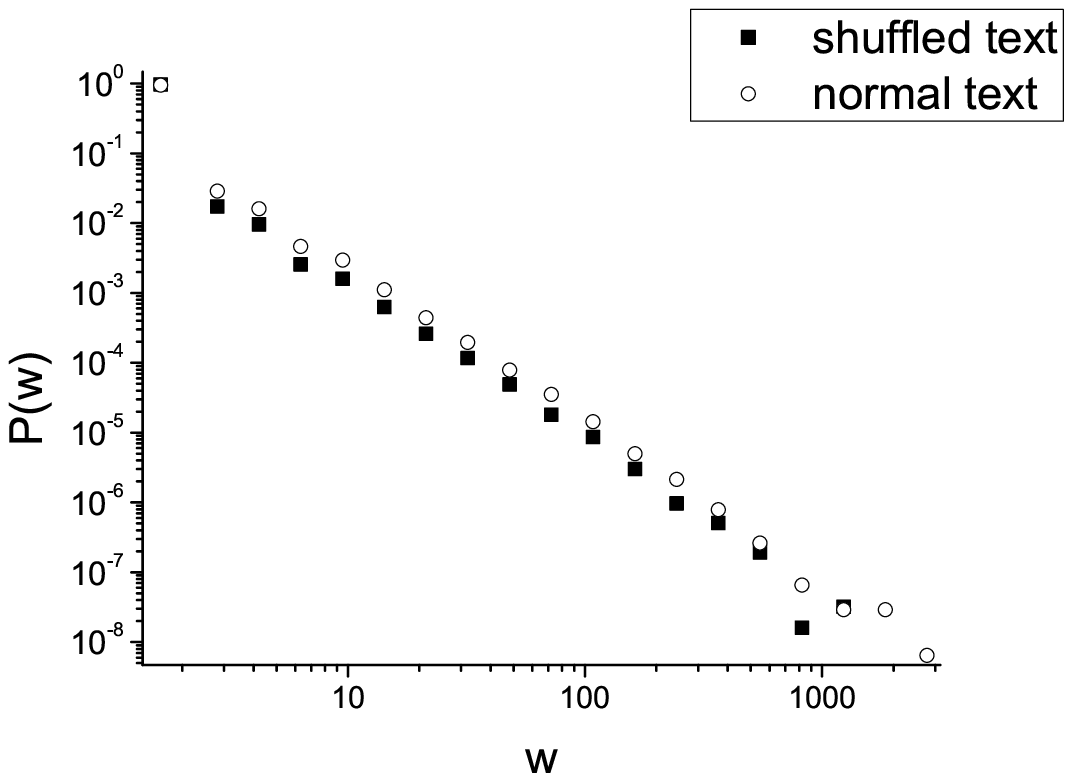}
         \includegraphics[width=0.49\textwidth]{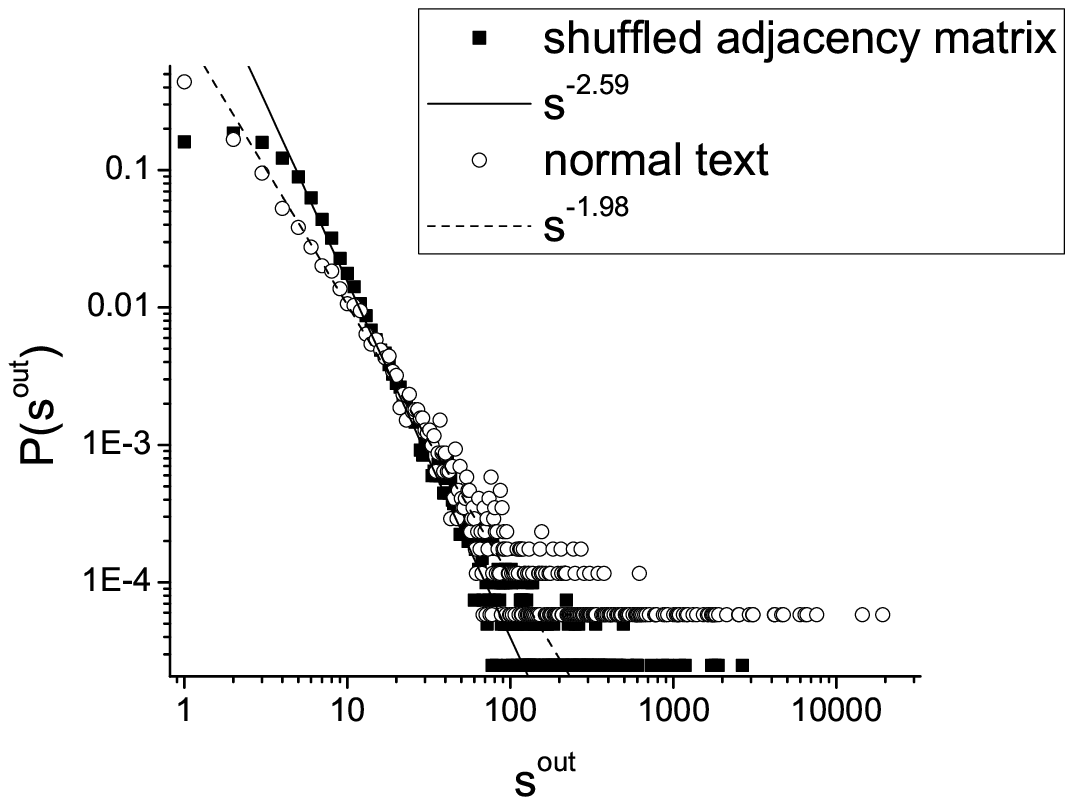}

\caption{\label{f3} Left panel: the weight distribution of \textit{Moby Dick} compared to the distribution obtained after shuffling the tokens of the novel, preserving the strength of the vertices. The result suggests that the scale free weight distribution is determined by the scale free strength distribution. Right panel: comparison between the out-strength distributions obtained from the real network and from the network obtained by shuffling the adjacency matrix preserving the weight distribution of the network. The result implies that the scale free strength distribution is not determined by the scale free weight distribution.}
\end{figure}

From the top of Fig.\ref{f1} we can see that the distribution of the degree of the vertices is not significantly altered in the shuffling process.  This result implies that the scale free degree distribution is induced by the scale free strength distribution. In fact the shuffling operation  redistributes the numbers that  occupy the rows of the weighted adjacency matrix in an uncorrelated way. This operation preserves the strength of the vertices, but changes their degree. However, as we can see, the average degree distribution is statistically preserved. This is an important result in network theory. In fact the distribution of the degree of a vertex, that is the distribution for the number of its nearest neighbours, is supposed to give a great deal of information about the system. In this case we can see that this measure cannot distinguish between a masterpiece and a meaningless collection of words.

To prove that the reverse doesn't hold, that is the scale free degree distribution doesn't induce a scale free strength distribution, we shuffle the network preserving the degree of the vertices. This operation consists in randomly redistributing the weights of the links between all the existing links. In reality it would consist in rewriting $Moby$ $Dick$ with the same vocabulary, the same total number of tokens, but changing the relative frequency of the tokens. We show the resulting strength distribution in the bottom of Fig.\ref{f1} compared to the one obtained from the real network. It is evident that the distributions obtained with the shuffling operation are peaked and well distinguishable from the ones obtained from the real network. The tails of these distributions are still power laws, but with a slope different from the one for the original network.  We derived the exponents for the tails of the distributions for $s\geq10$ with the \emph{method of maximum likelihood} proposed by Newman in \cite{pl} and we calculated the error on the exponent with the \emph{bootstrap method} \cite{bs}, with 2000 replicas for the normal network and 4000 replicas for the shuffled network. In this way we found that the real network strength distribution displays a power law tail with exponent $-1.98\pm 0.02$, while the shuffled network strength distribution displays a power law tail with exponent $-2.19\pm 0.01$ for the in-strength and $-2.17\pm0.01$ for the out-strength distribution.

The straight connection between strength distribution and degree distribution can be explained by the power law relation between strength and degree. In Fig.\ref{f0} we show the empirical data for $<s(k_{out})>$ before and after the shuffling process. In both the cases we observe a relation of the type $<s(k)>\propto k^\delta$. Since $P(s)\propto s^{-\gamma}$, then $P(k)\propto k^{\delta(1-\gamma)-1}$. This power law relation holds only in average though. In fact, as we'll discuss at the end of this section, the relation between strength and degree is more complex and deep.

The weight distribution $P(w_{ij})$ for linguistic networks, such as that for other scale free weighted networks, is a power law. From the left panel of Fig.\ref{f3} we can see that if we shuffle the text the resulting weight distribution doesn't vary significantly. The only difference that can be appreciated between the distributions of Fig.\ref{f3} is in their tails. In fact the range of values for the weight  in the normal text is larger than that in the shuffled text. If we shuffled the entries of the adjacency matrix without preserving the strength of the vertices, the weight distribution would dramatically change in a uniform distribution. Again we can argue that the power law distribution for the weight of the vertices is induced by the scale free strength distribution.

To prove that the strength distribution is not implied by the weight distribution we shuffle the network preserving its weight distribution. This is done by randomly shuffling the cells of the adjacency matrix, preserving the weights of the shuffled cells. In the right panel of Fig.\ref{f3} we show the results of this experiment. For the shuffled network the resulting strength distribution is peaked. Nevertheless it displays a power law behaviour for many decades, but with a slope much steeper than that of the real network. Again we derived the exponents for the tails of the distributions for $s\geq10$ with the method of maximum likelihood  and we calculated the error on the exponent with the bootstrap method using 5000 replicas for the shuffled network. In this way we found that the  shuffled network out-strength distribution displays a power law tail with exponent $-2.59\pm 0.02$, compared to the exponent of the tail of the original distribution, that is $-1.98\pm 0.02$. Moreover the fat tail is much shorter than that one for the real network.

\begin{figure}[!htbp]\center
         \includegraphics[width=0.49\textwidth]{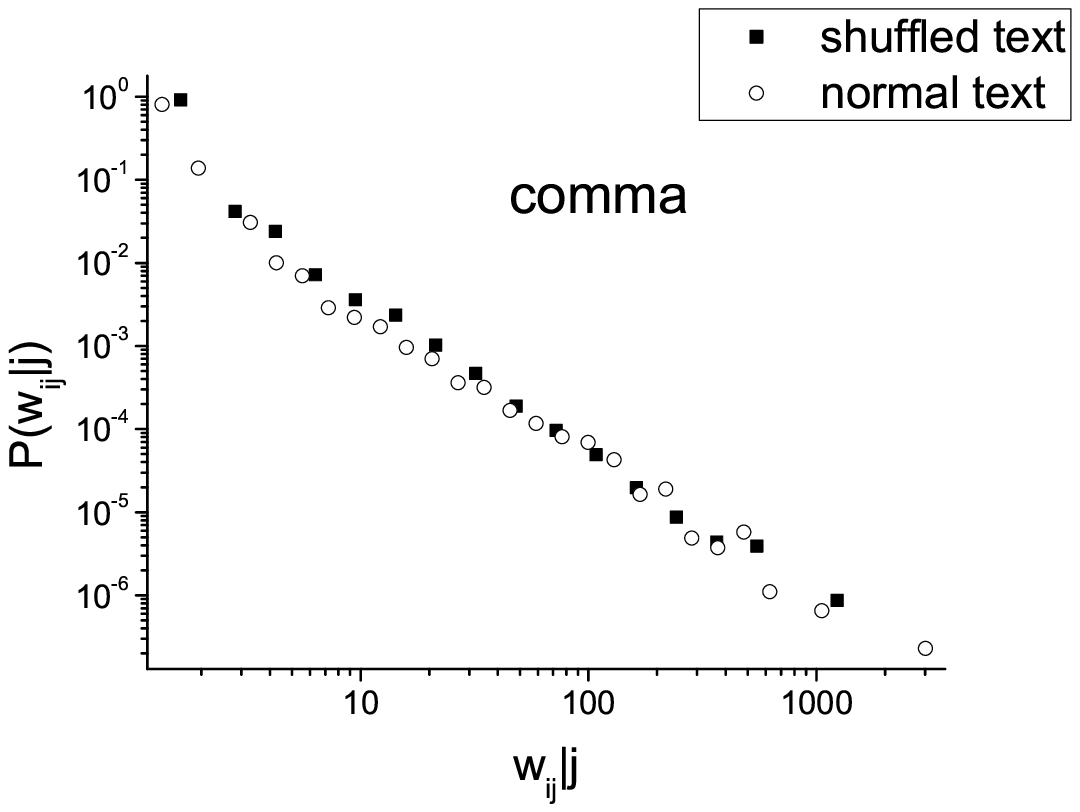}
          \includegraphics[width=0.49\textwidth]{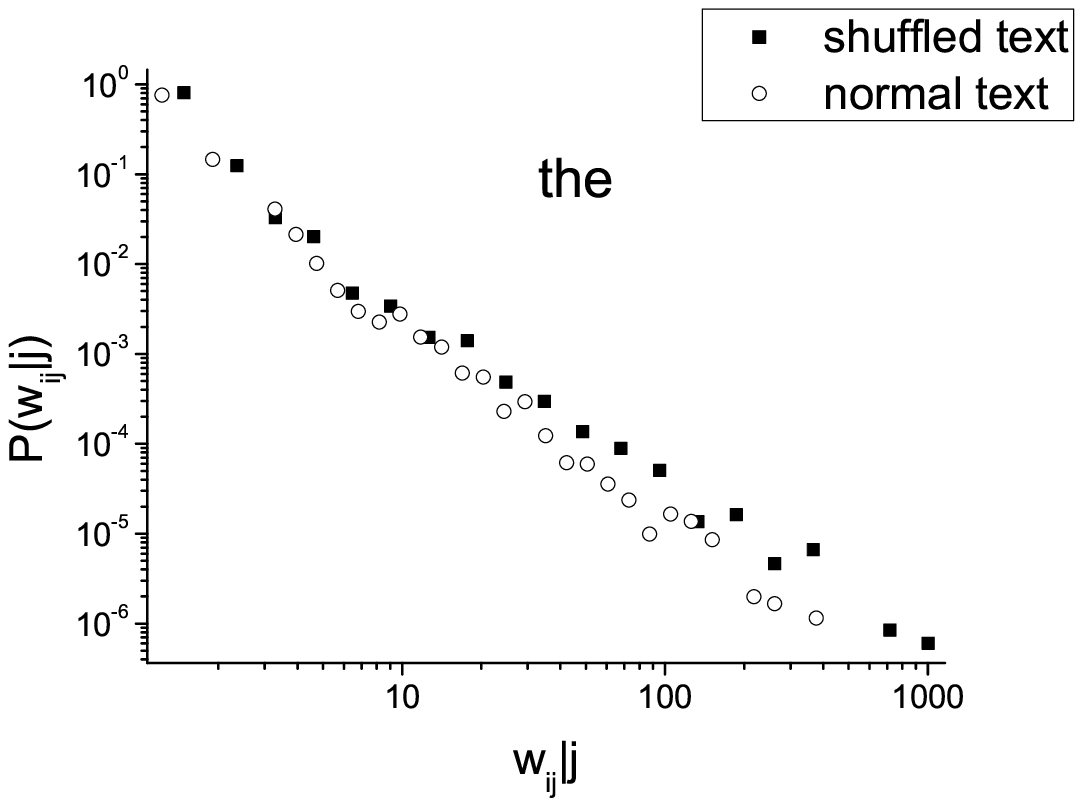}
           \includegraphics[width=0.49\textwidth]{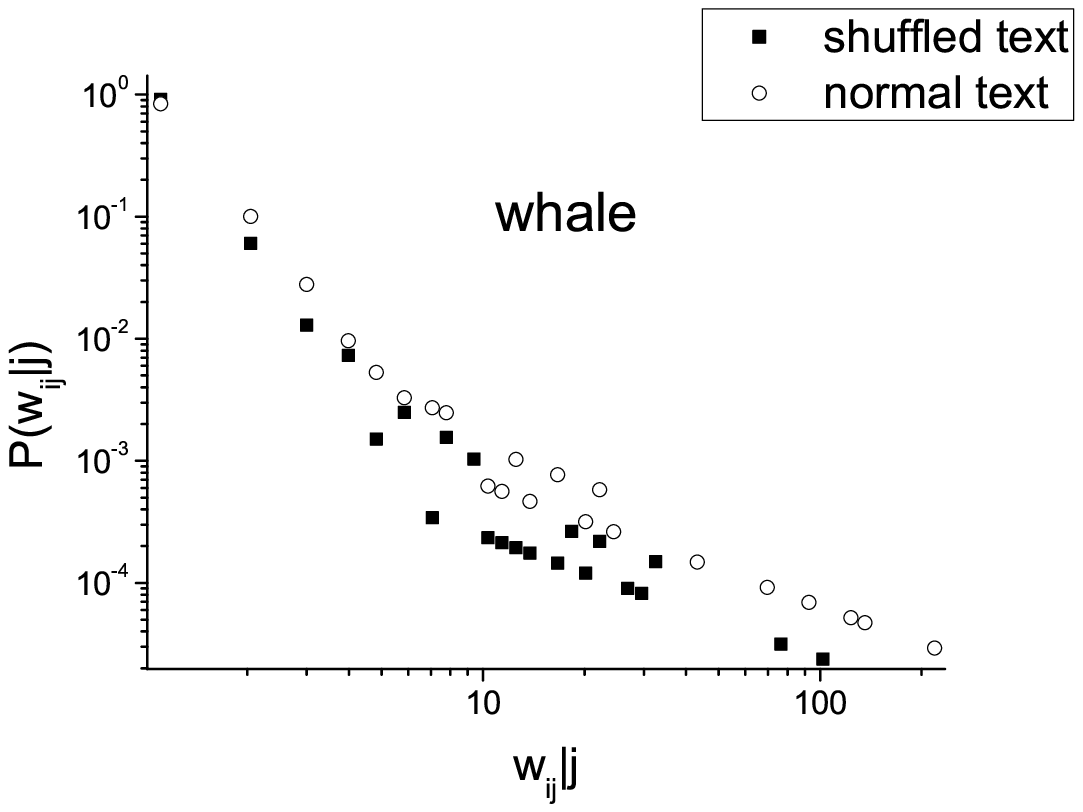}
            \includegraphics[width=0.49\textwidth]{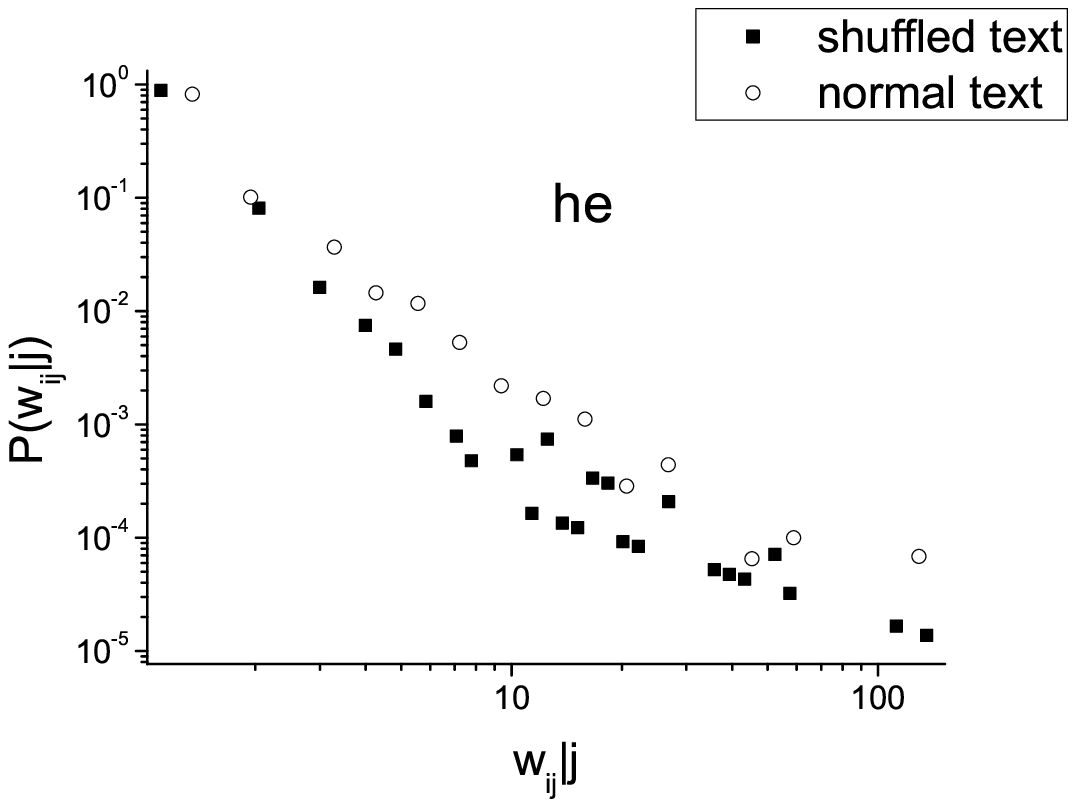}

\caption{\label{f4} Distribution of the weights of the links of a single token $P(w_{ij}|j)$ in \textit{Moby Dick} compared to the distribution obtained after shuffling the tokens of the novel, preserving the strength of the vertices for the tokens $comma$, $the$, $whale$ and $he$.}
\end{figure}

A measure that is interesting to study in a weighted network is the distribution of the weights of the links of a single token, that is $P(w_{ij}|j)$. In Fig.\ref{f4} we show the distributions of $P(w_{ij}|j)$ arising from the normal and the shuffled text for four very frequent tokens. Even in this case no real differences emerge.

\begin{figure}[!htbp]\center
        \includegraphics[width=0.49\textwidth]{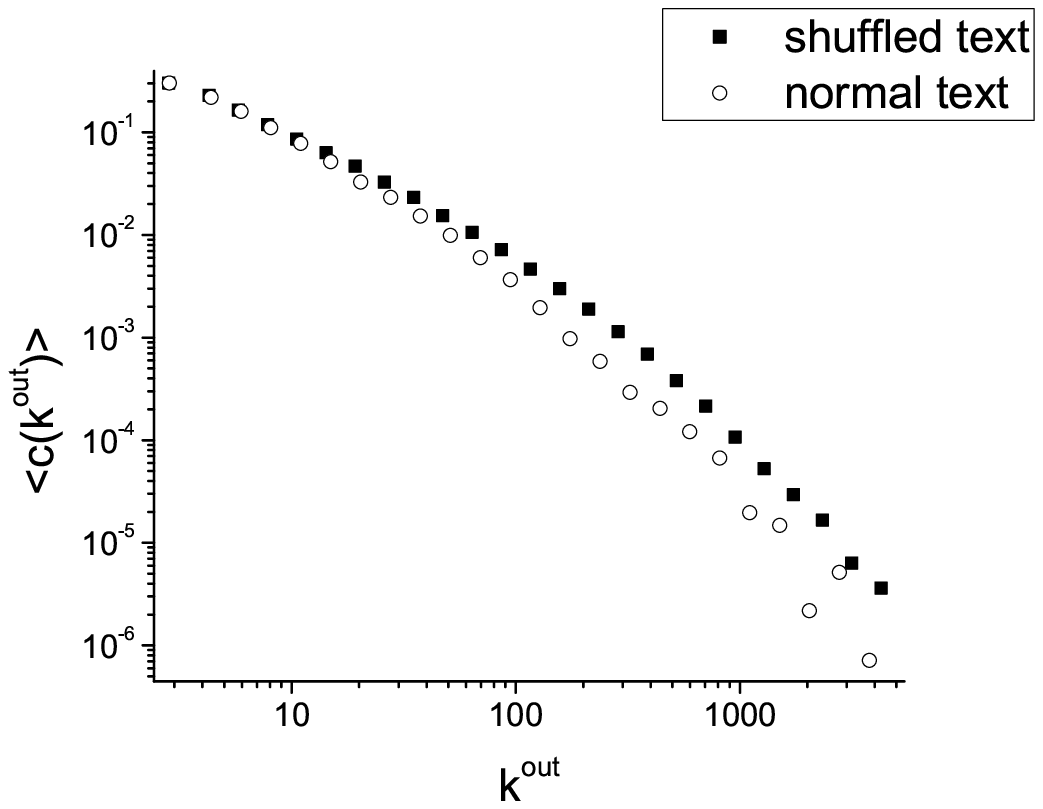}
         \includegraphics[width=0.49\textwidth]{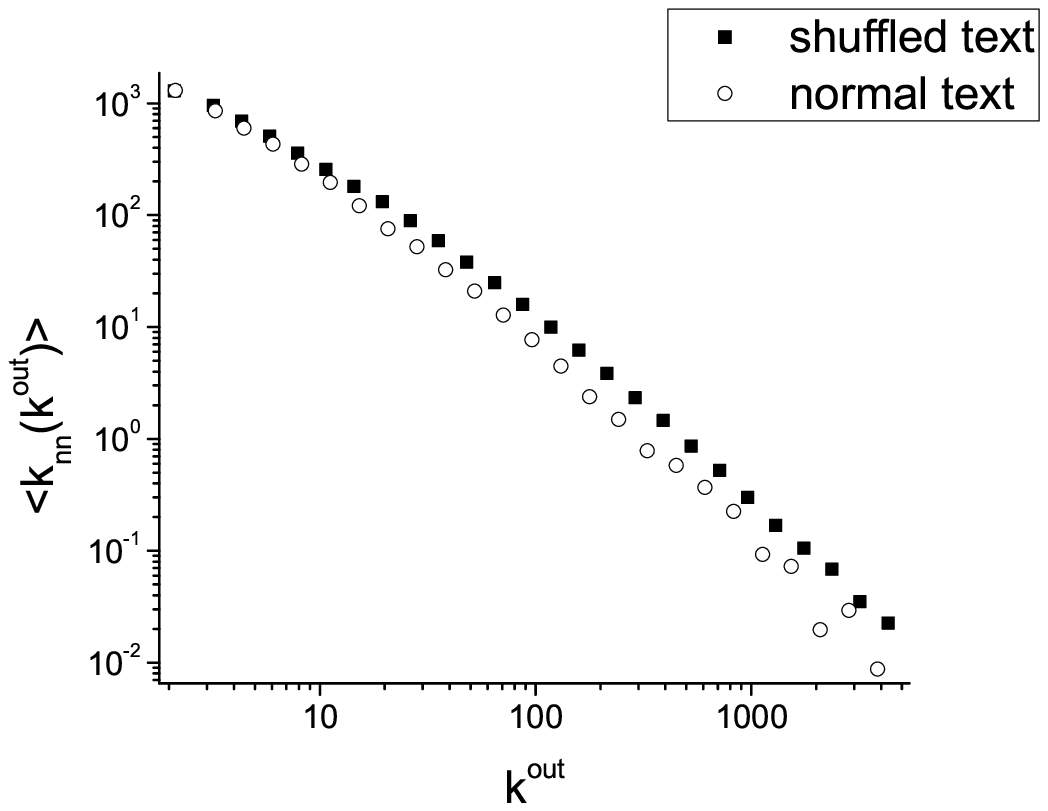}

\caption{\label{f2} In the left panel: average clustering coefficient $<c(k^{out})>$ of \textit{Moby Dick} compared to the one obtained after  shuffling the tokens of the novel, preserving the strength of the vertices. In the right panel: out-nearest neighbour degree distribution of \textit{Moby Dick} compared to the distribution obtained after  shuffling the tokens of the novel, preserving the strength of the vertices.}
\end{figure}

A measure that is always considered to characterise the topology of a network is the clustering coefficient $c(k)$, that measures the number of nearest neighbours of a vertex with degree $k$ that are interconnected, divided by the maximum allowed number of such interconnections. In a directed network the classical formula for $c(k_i)$ \cite{1}, for vertex $i$ with degree $k_i$, has to be changed in $c(k_i)\equiv\frac{d_i}{k_i(k_i-1)}$ \cite{12}, with $k_i>1$, where $d_i$ is the number  of directed links between the nearest neighbours of vertex $i$. In the left panel of Fig.\ref{f2} we show the measured $<c(k^{out})>$ for the normal and the shuffled text.
No big differences emerge. In both  cases the highest average clustering coefficients are associated with vertices with small degree and in both the cases we have a double slope power law decay. Nevertheless we have to mention that, as already noted in \cite{12}, when we decrease the size of the bins,   the real text data shows a more structured texture than the smoothed data of the shuffled text. The smoothed curve is a signature of a stochastic process,while the complexity of language allows for more complicated behaviour, creativity for instance, that is not stochastic.

To study second order correlations   we have to deal with the nearest neighbour average degree $k_{nn}$. In the right panel of Fig.\ref{f2} we show the average nearest neighbour out-degree as a function of the out-degree for the normal and the shuffled text. Even   in this case no striking differences emerge, the disassortative feature of the network remains unchanged in both cases. As for the clustering coefficient, if we decrease the size of the bins, the real text data shows a more structured texture than the smoothed data of the shuffled text \cite{12}. This sort of behaviour has been found recently  in other information networks \cite{17}.

\begin{figure}[!htbp]\center
         \includegraphics[width=0.49\textwidth]{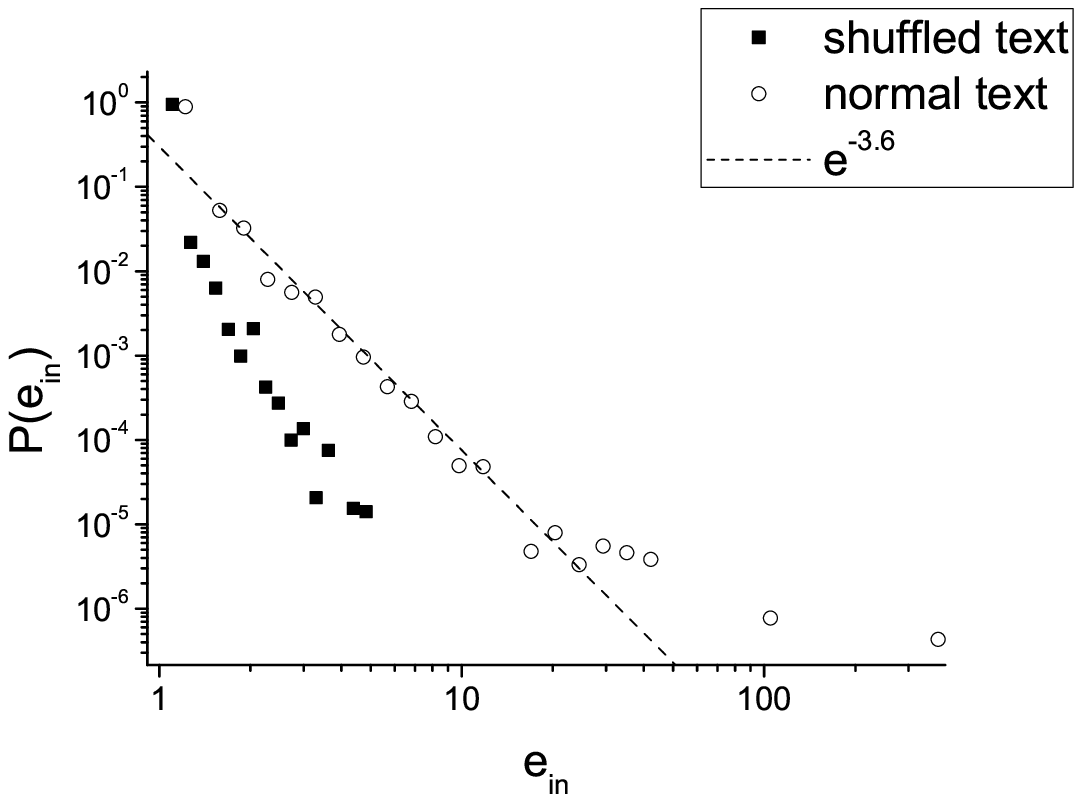}
          \includegraphics[width=0.49\textwidth]{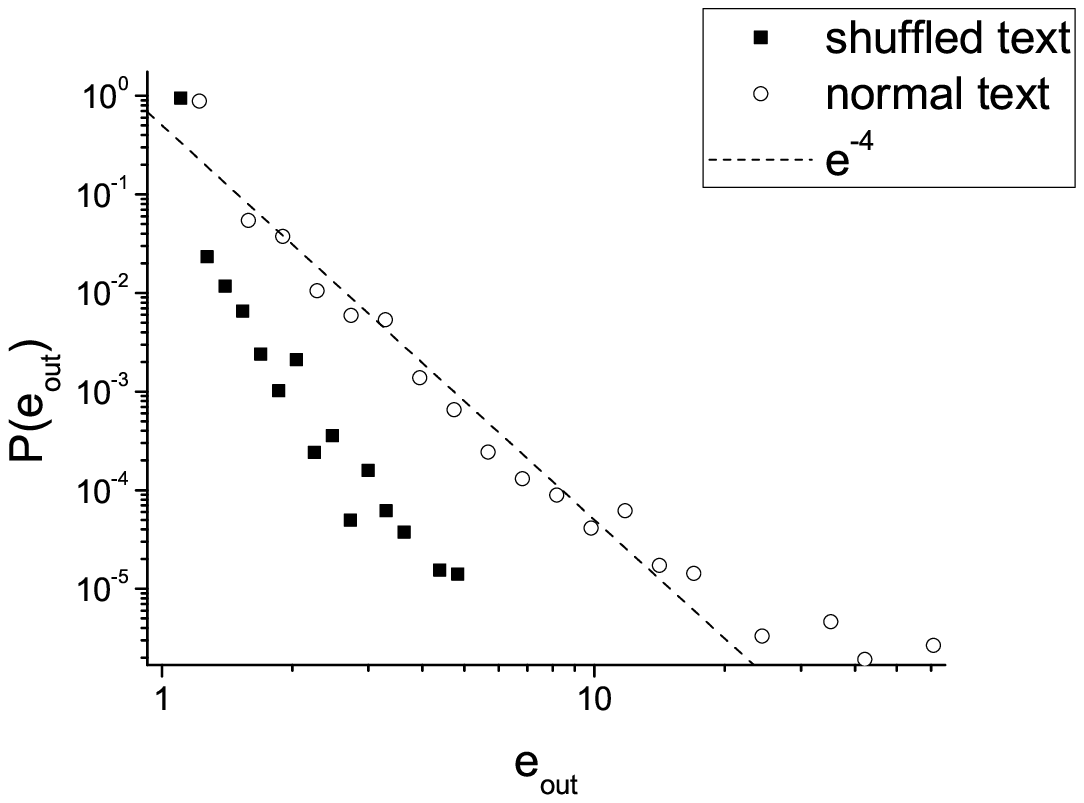}

\caption{\label{f5}The out and in-selectivity distributions of \textit{Moby Dick} compared to the distributions obtained after  shuffling the tokens of the novel, preserving the strength of the vertices. It is shown that this measure can effectively distinguish the shuffled text from the real one. }
\end{figure}

Considering the average number of out and in-links per connection each node has, we finally find a measure that can distinguish the shuffled text from the real one. We then define for the vertex $i$ the $out$ and $in-selectivity$ as
\begin{equation}
 e^{out/in}_i\equiv\frac{s_i^{out/in}}{k^{out/in}_i}=\frac{s_i}{2k^{out/in}_i},
 \end{equation}

  the last equality holding for Eulerian networks, so that $e\geq 1$.  The selectivity is a measure that can capture  the effective distribution of numbers in the weighted adjacency matrix. To understand its meaning we can consider that the token with biggest out-selectivity in $Moby$ $Dick$ is $``Mr"$, with $e^{out}_{Mr}=63$, that is the token $``Mr"$ is always followed by the token $``dot"$. Then, in descending order, we find $``Moby"$, $``didn"$, $``won"$, $``instead"$, etc... that are all tokens very  selective in the choice of their out-neighbourhood and that form the so called \textit{morphologic structures} of the language \cite{14}. Most of the tokens with small out-selectivity  are tokens that appear just a few times in the text
(core lexicon\cite{13}), but  there are also tokens that appear many times in the text and that are  characterised by small values of the out-selectivity. These are the tokens that connect with a different token each time, that is they are not selective in the choice of their neighbourhood. For this latter case words like $``really"$, $``strangely"$, $``grow"$, $``real"$, $``terrible"$,etc...have out-selectivity $e^{out}=1$. The in-selectivity has the same meaning as the out selectivity, but probing at the in-neighbourhood. The token with biggest in-selectivity in $Moby$ $Dick$ is $``s"$, with $e^{in}_s=360.4$, that is almost preceded by the token $``'"$. Then we have $``ll"$, $``em"$, $``Dick"$, etc...

In Fig.\ref{f5} we show the distribution $P(e^{out/in})$ for the out and the in-selectivity of the vertices compared to the same distributions obtained for the shuffled text.  An important aspect for the vertex selectivity measure is its range. For the normal text the range for the vertex selectivity is much larger than the one for the shuffled text, and, in the case of the in-selectivity, this difference is of one order of magnitude. This comes from the fact that in the real text the tokens are  selective in choosing their neighbours and form very specialised local structures. The lack of those local structures determines the small values for the selectivity in the shuffled text and so the big difference of the selectivity distributions between the shuffled and the real text. Another important point to notice in Fig.\ref{f5} is that in the case of the real text the distribution of the selectivity appears to follow a power law for many decades of the selectivity.

\begin{figure}[!htbp]\center
         \includegraphics[width=0.49\textwidth]{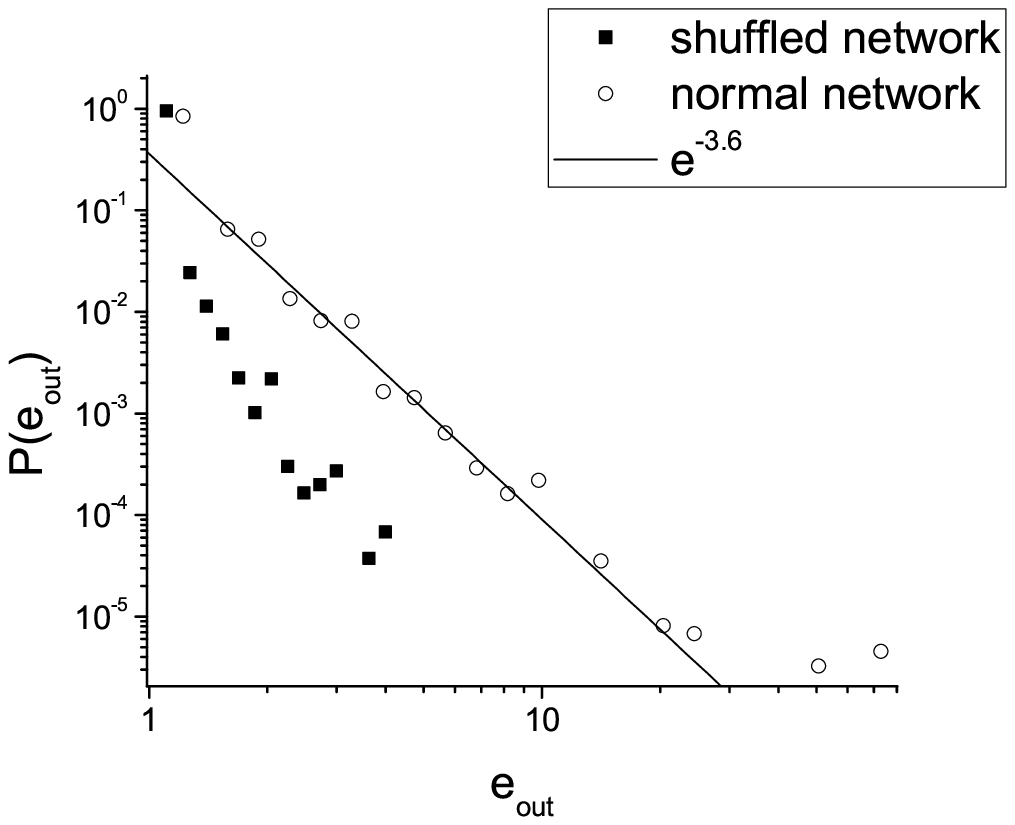}
          \includegraphics[width=0.49\textwidth]{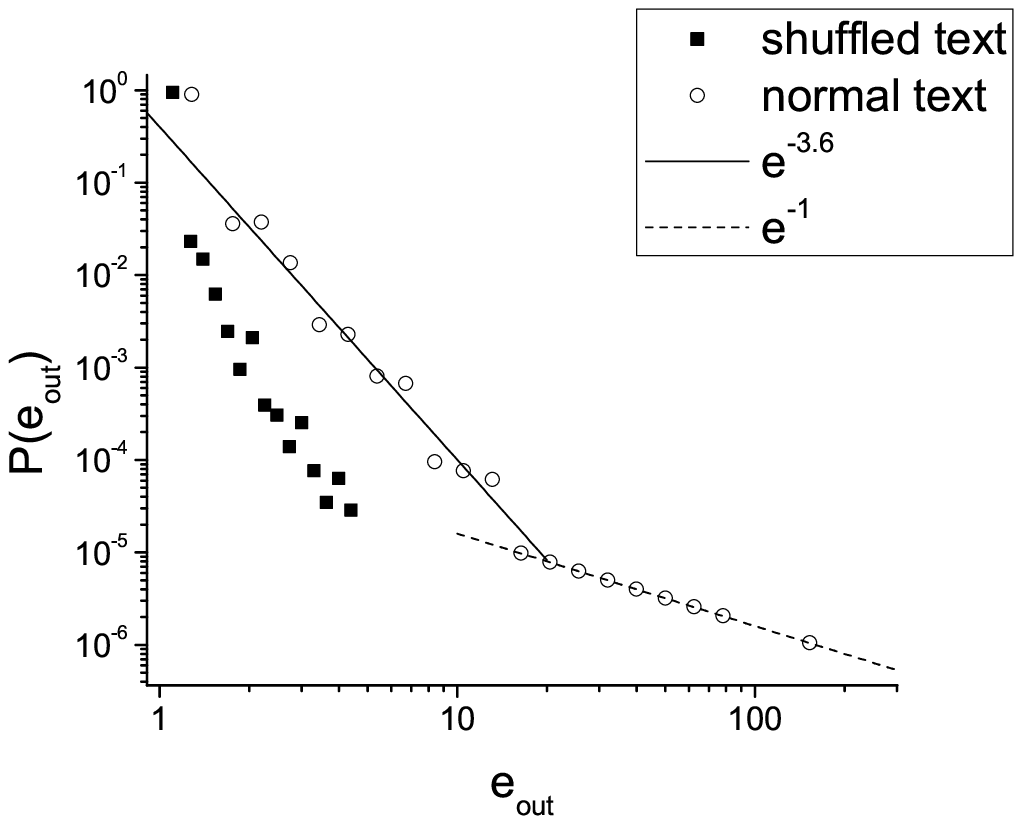}

\caption{\label{fn}Out-selectivity distributions of \textit{Nineteen Eighty-four} on the left and \emph{On the road} on the right compared to the distributions obtained after  shuffling the tokens of the novel, preserving the strength of the vertices. }
\end{figure}

A question one could raise is about the robustness of selectivity, that is if its regularities are present in other linguistic networks. To answer this question we consider the same analysis on two other novels,  \emph{``Nineteen Eighty-four"} by George Orwell \cite{hl6}, and \emph{``On the road"} by Jack Kerouac \cite{hl32}. In Fig.\ref{fn} we show the out-selectivity distribution for the two novels compared to the ones obtained after shuffling the networks, preserving the strength of the vertices. We find the same behaviour found in \textit{Moby Dick}. The distributions appear to follow a power law decay with exponent around $-3.6$ for many decades of the out-selectivity $e_{out}$. The range of the out-selectivity for the real texts is much larger then the one found in the shuffled texts. Moreover in the case of \textit{Moby Dick} and  \emph{Nineteen Eighty-four} the out-selectivity distribution  is unresolved for large values of the out-selectivity, maybe because of finite size effects. In the case of \emph{On the road} (right panel of Fig.\ref{fn}) the distribution tail for large values of out-selectivity is cleaner and  is very well fitted by a power law with exponent $-1$. It is important to notice that just the $0.001\%$ of  the very frequent tokens of the novel generates this second tail.

We believe that, since the vertex selectivity is related to the morphologic structures of language, its behaviour is strictly related to the style of the writer. Anyway striking similarities are evident in its global statistical behaviour in the three novels considered.

\section{Scientific collaboration network}

\begin{figure}[!htbp]\center
         \includegraphics[width=0.49\textwidth]{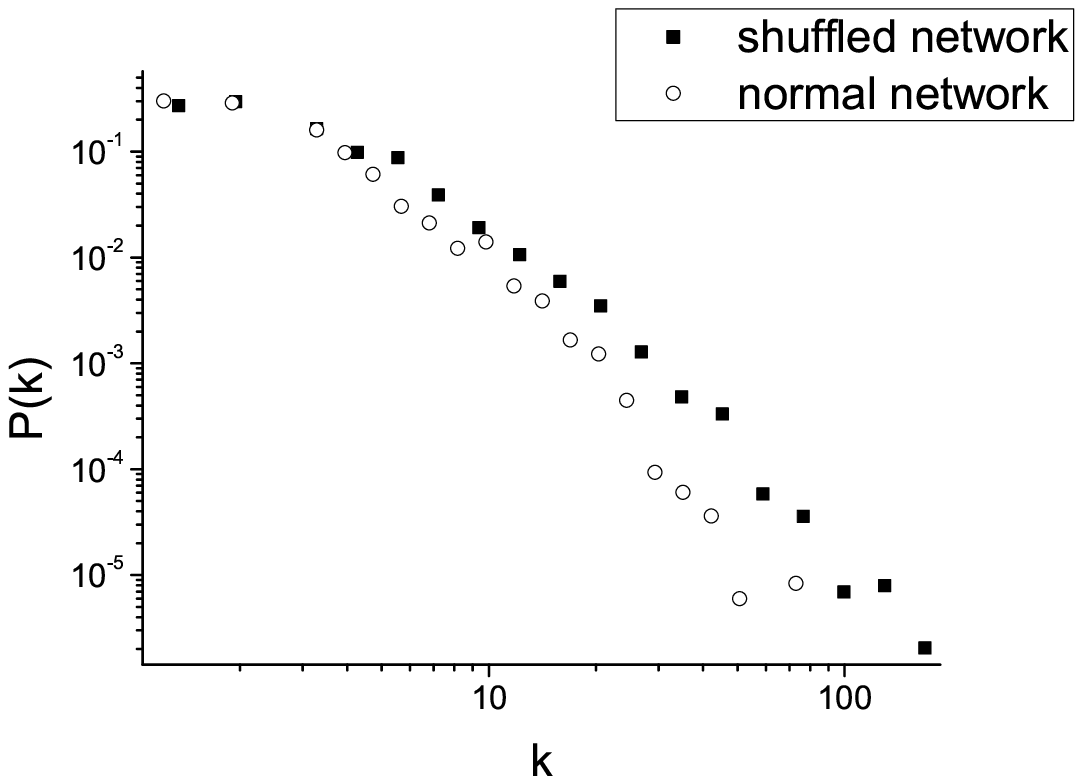}
          \includegraphics[width=0.49\textwidth]{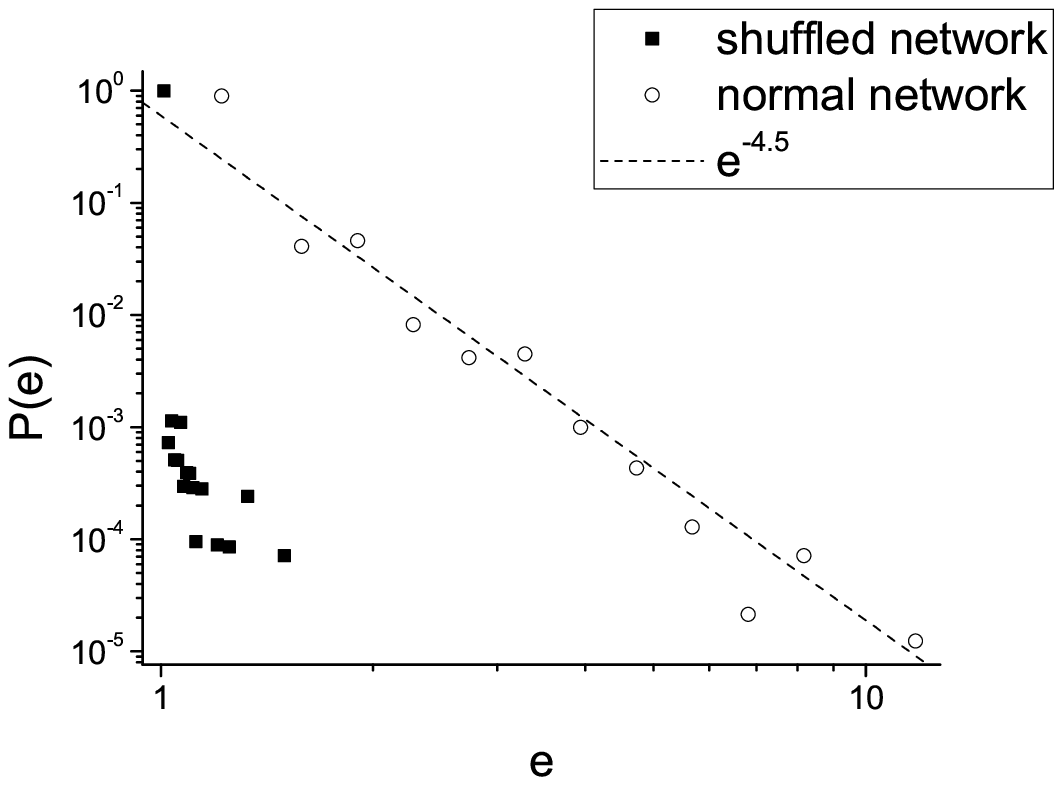}

\caption{\label{f6} In the left panel the degree distribution for the scientific collaboration network compared to the distribution obtained after shuffling the tokens of the network, preserving the strength of the vertices. In the right panel the selectivity distribution of the scientific collaboration network compared to the distribution obtained after the shuffling process.}
\end{figure}

To test our results and to understand if they are suitable for the study of other types of networks that can be analysed in a weighted framework, we consider a scientific collaboration network. This kind of network, showing scale-free properties, has been analysed by a number of authors \cite{15,16}. We consider the  scientific papers published between the 2000 and 2007 containing the word $network$ or $networks$ in the title. We define the vertices of the network to be the different authors of the paper and two vertices to be connected if they represent coauthors of the same paper. The weights of the links are defined as the number of times two authors coauthored. In this case, since the relation of coauthorship is reflexive, the network is undirected and the weighted adjacency matrix symmetric. The selectivity for vertex $i$ is then defined as $e_i\equiv \frac{s_i}{k_i}\geq1$.  The resulting network, obtained by 5335 papers, has 9503 vertices and 26966  undirected links with an average strength $<s>\approx5.67$, average degree $<k>\approx 4.57$ and average selectivity $<e>\approx1.18$.

We then shuffled the vertices of the network and  considered the resulting network. Since the strength for a vertex represents the number of papers an author wrote, the shuffling operation, as in the previous case, doesn't alter the strength distribution of the network. In the left panel of Fig.\ref{f6} we show the comparison of the degree distributions between the shuffled and the real network. We don't have to be alarmed by the fact the scale free behaviour in this case is not as straight as in the language network, since this is a characteristic of this peculiar network (for instance see analysis in \cite{15}). The important aspect of the analysis is that it is very difficult to infer which of the networks is the shuffled one by looking at its degree distribution. It is possible to observe that in the shuffled case the range of values for the degree is larger. This is implied by the lack of topological correlations of the shuffled system, so that the vertices tend to link to a larger number of different vertices. Nevertheless this difference is so slight that it is not possible to consider it as a way of discriminating between the two different networks. In particular, in the case of the linguistic network, the maximum degree of the real network is  larger then the one found in the shuffled network, even if the average degree is smaller. Then we can say again that the scale free degree distribution is implied by the scale free strength distribution.

On the right panel of the same figure we show a comparison between the distributions of the selectivity for the two networks. This time the difference is striking, the normal network displaying a power law distribution with exponent around -4.5, the shuffled network displaying an ill-defined distribution.
In this case large values of the selectivity characterise  authors that have exclusive relations with a few other authors. So, for instance, an author having a great affinity with another author,  that is an author who had written all or almost all his/her papers with the same coauthor will be characterised by a large value of selectivity. Authors who have published just one paper or who have published all their papers with different coauthors will be characterised by selectivity equal to 1.

\begin{figure}[!htbp]\center
         \includegraphics[width=0.49\textwidth]{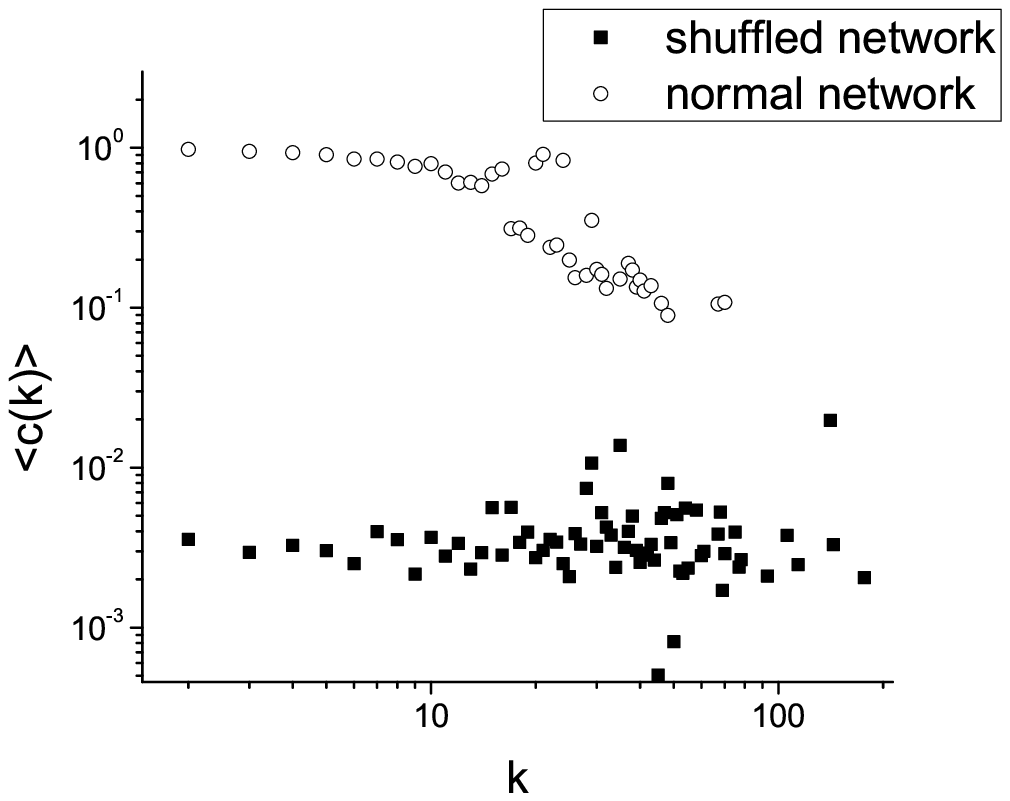}
          \includegraphics[width=0.49\textwidth]{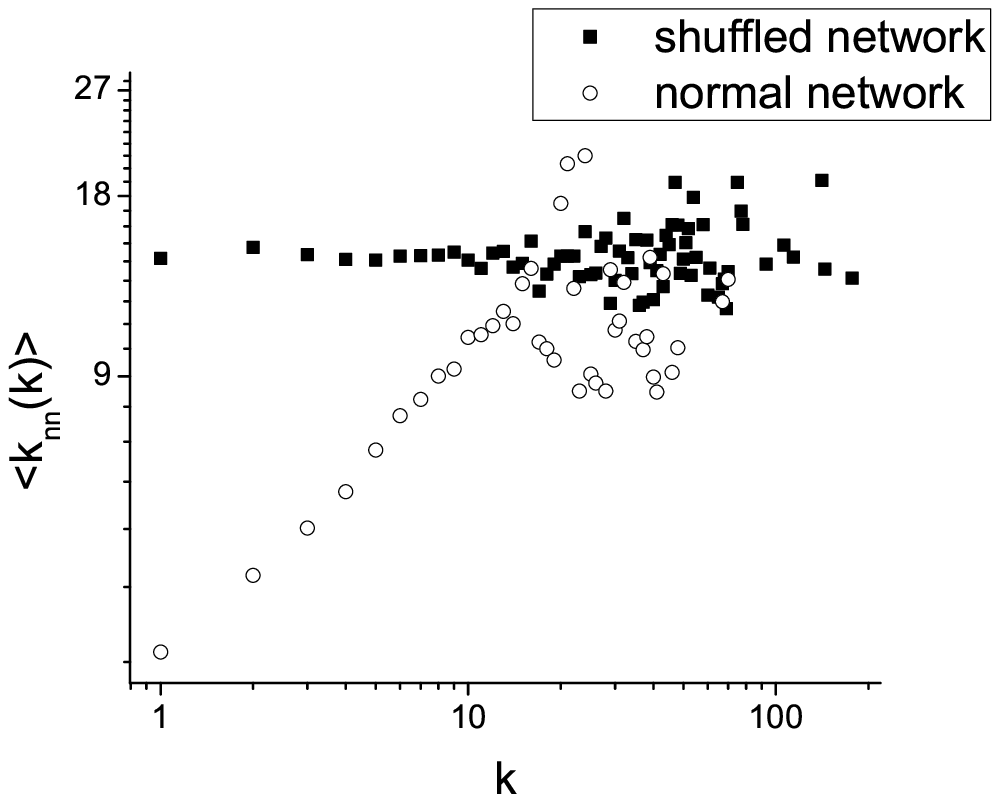}

\caption{\label{f8} In the left panel the average clustering coefficient distribution $<c(k)>$ of the scientific collaboration network compared to the distribution obtained after  shuffling the tokens of the network, preserving the strength of the vertices. In the right panel the average nearest neighbour degree distribution $<k_{nn}(k)>$ of the scientific collaboration network compared to the distribution obtained after  shuffling the tokens of the network, preserving the strength of the vertices.}
\end{figure}

Interestingly,  the clustering coefficient and the nearest neighbour degree distributions  for the coauthorship network show a very different behaviour from their shuffled counterparts. In the left panel of fig.\ref{f8} we show the comparison of the average clustering coefficient $<c(k)>$, averaged for each degree $k$, for the normal and the shuffled network. For the normal network we can observe a highly clustered structure, especially for vertices with small degree. This behaviour is due to the fact that all the papers written by more then two authors form clustered structures. In the shuffled version of the network those triangular structures completely disintegrate.  In the right panel of fig.\ref{f8} we show the comparison of the average nearest neighbour degree $<k_{nn}(k)>$, averaged for each degree $k$, for the normal and the shuffled network. In the case of the real network, an assortative behaviour is evident for small values of the degree, after that the behaviour appears to fluctuate around an average. In the case of the shuffled network the assortative behaviour completely disappears and the vertices don't show any preference in the way they connect.  We conclude that those measures catch the distinctive trait of the network, but at the same time we have to ask ourselves to which networks these measures  can be applied to.

\section{A stochastic model}

In this section we want to show that the selectivity measure can detect the local structures of weighted networks. For this reason we introduce an ad-hoc growing network, whose growth is based on both global (hereafter GPA) and local preferential attachment (hereafter LPA). We want to reproduce the topological behaviour of the coauthorship network investigated in the previous section. In fact the LPA \cite{12} enforces the existing network edges, creating preferential and differential local structures, and enhancing second order correlations.

We start with a network of 50 pairs of connected vertices. Then, at each time step, we introduce a new vertex in the network and we connect it to an old vertex via GPA, that is with probability $\Pi$ proportional to the strength of the old vertices, $\Pi=\frac{s_i}{\sum_js_j}$. We then add $m=2$ new edges in the network. With probability $p$ one of the new edges will connect to vertices $i$ and $j$ chosen via GPA, that is with probability proportional to
the product of the strengths of the two vertices, $\Pi=\frac{s_is_j}{\sum_{k,l}s_ks_l}$. With probability $1-p$ one of the new edges will connect two already connected vertices via LPA, that is with probability proportional to the weight of the old edges, $\Pi=\frac{w_{ij}}{\sum_{i,j}w_{ij}}$.

The 50 initial pairs of connected vertices represent 50 pairs of authors writing  about a new scientific topic. The new incoming vertex, connected via GPA, represents a new scientist joining the community and attracted to coauthor with authors who have already written many papers on the topic. The new edges introduced with probability $p$ via GPA represent the mixing inside the scientific community, encouraged by the popularity of the authors. The new edges introduced with probability $1-p$ via LPA represent coauthors that carry on writing together. The value of $m$ is chosen to obtain an average strength of 6. The simulation was run to obtain a network of 10000 vertices and 30000 edges.

\begin{figure}[!htbp]\center
         \includegraphics[width=0.49\textwidth]{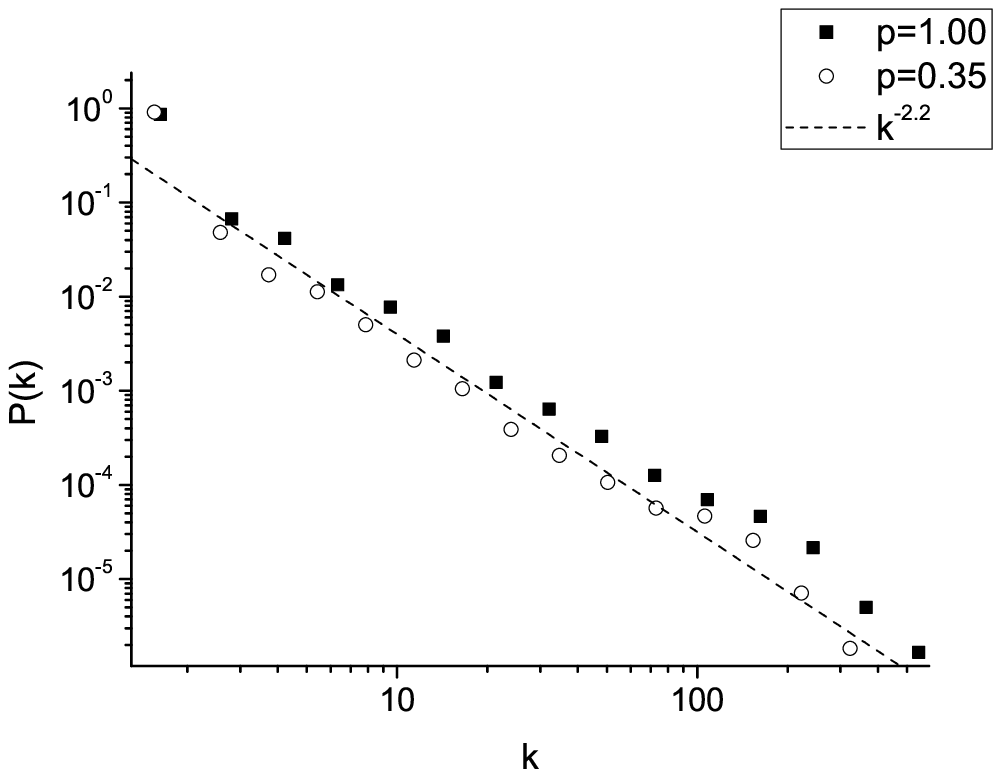}
          \includegraphics[width=0.49\textwidth]{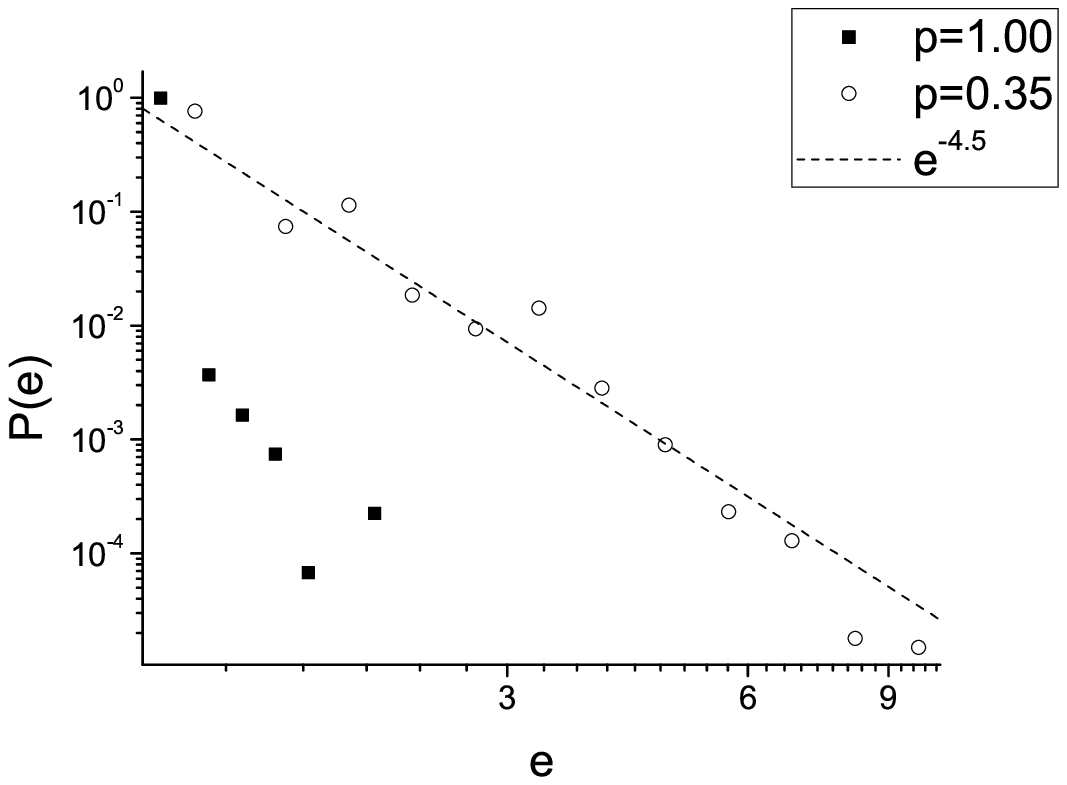}
            \includegraphics[width=0.49\textwidth]{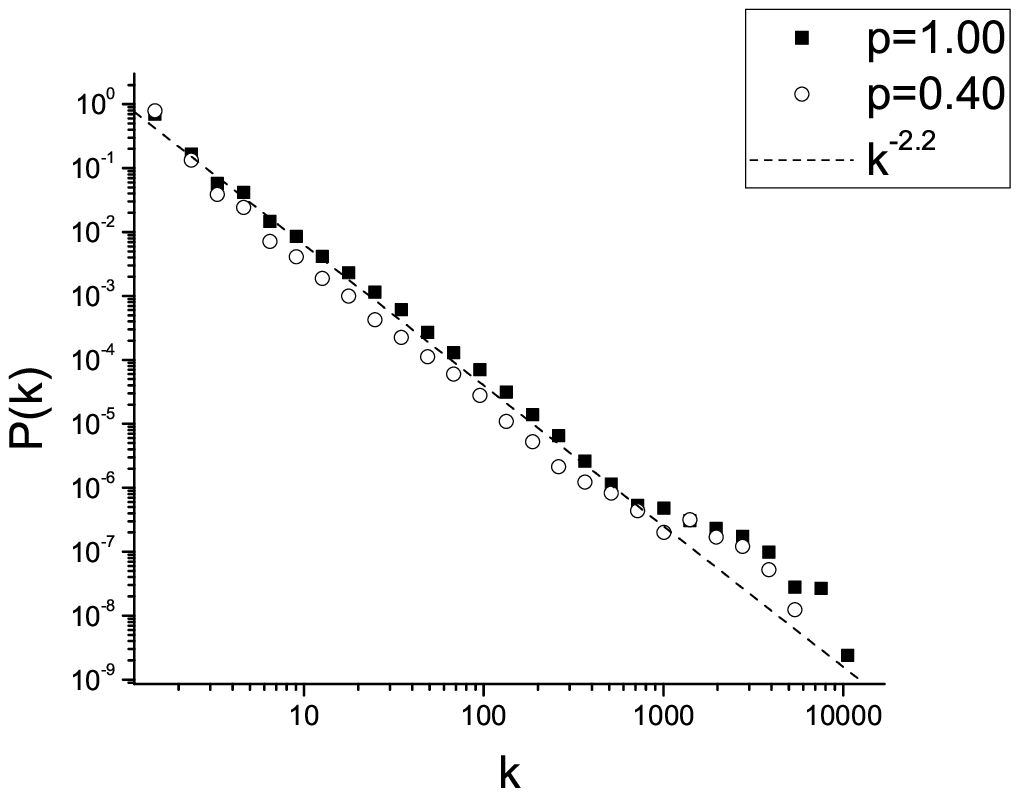}
          \includegraphics[width=0.49\textwidth]{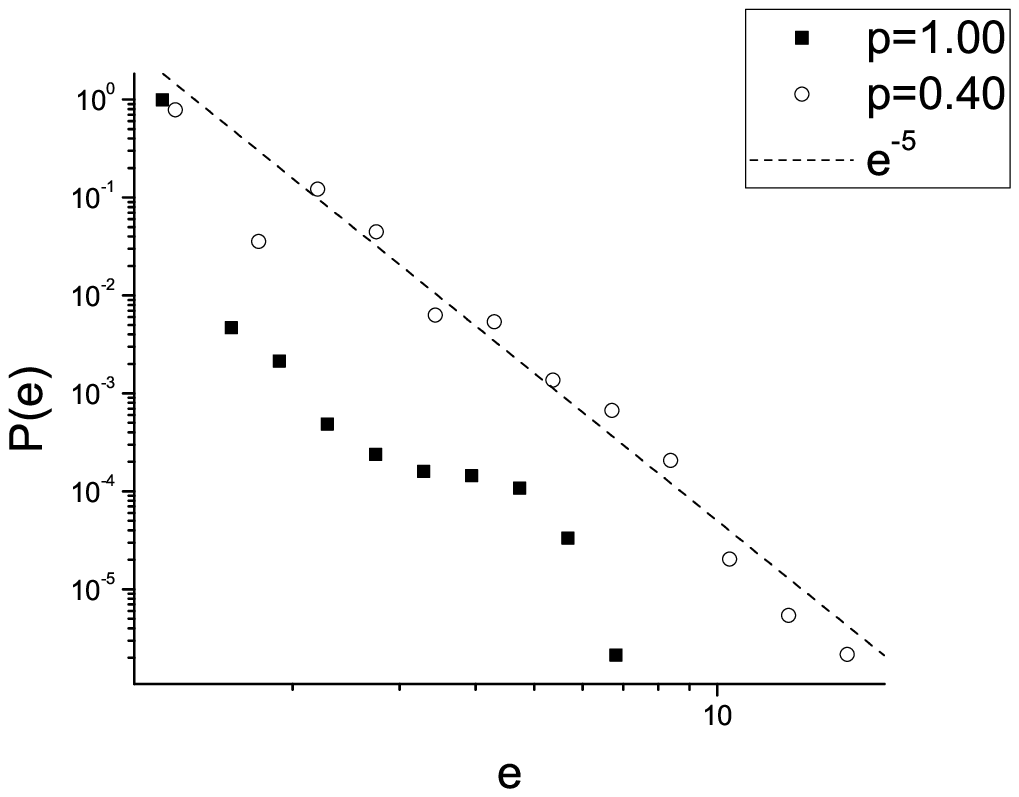}
\caption{\label{f7} Top panels: in the left panel the degree distributions of the stochastic network for different percentages $(1-p)$ of LPA during the growth process. No relevant differences exist between the distributions. In the right panel the selectivity distributions of the same networks. Significant differences emerge. We find the same slope for the power law found in the empirical behaviour of Fig.\ref{f6} for $p=0.35$. Bottom panels: the previous simulation for a network of $10^5$ vertices and $10^6$ links for different values of $p$. Again no relevant differences emerge for the degree distribution (left panel), while the selectivity distribution (right panel) is highly affected by the artificial correlations induced by the LPA.}
\end{figure}

In the top panel of Fig.\ref{f7} we show the analysis of the resulting network for different values of $p$. In particular we compare the results for the network obtained with $p=1$, that is a total GPA, and the one obtained with $p=0.35$, in which the LPA is the dominant growth  process. The scale free degree distribution for the two resulting networks has the same power law exponent, as shown in the left panel of Fig.\ref{f7}. From the right panel we can see that, while a full GPA attachment rule produces a distribution for the selectivity similar to the one measured in the shuffled networks of the previous sections, the mixed GPA and LPA attachment produces a selectivity distribution that fits with the empirical behavior of Fig.\ref{f6}.

To check if the behaviour of the model is stable for larger networks, we show in the bottom panels of Fig.\ref{f7} the same simulation for a network of $10^5$ vertices and $10^6$ edges. The  values considered for $p$ are $p=1$ and $p=0.4$. Again we can see from the left panel that the scale-free degree distribution is not altered by the local attachment processes. In the right panel we can observe that for the network grown by LPA the selectivity distribution follows approximatively a power law with exponent -5, while for the network grown without LPA the selectivity distribution is ill defined.

\section{Conclusions}

Many real networks in nature are weighted networks. These networks, for instance food webs, ecological networks, linguistic networks, social and urban networks deserve special attention, as their study will improve our knowledge of their organisation and allow us to understand the way to act on them. In this paper we have tried to address the question of the dependence between different properties of weighted networks. The complexity of such large real systems seldom allows an analytical approach to the problem. This is the reason we applied an algorithmic procedure that, even if  not as strong as the analytical one, is strong enough to yield conclusions, and is reproducible.  In the light of this, we considered the standard measures on networks and, after shuffling the networks conserving each time a peculiar symmetry of the system, we compared the new measures to understand which properties of the networks were preserved.

For this study we first considered a linguistic network. The reason for this choice is that a novel is a closed system that can be analysed in the framework of network theory without ambiguity and from many points of view, such as time series analysis and information theory analysis. Moreover the availability of data sets in linguistics is huge,  data sets can be as large as desired and we  know exactly the formal rules of the composition, that is the syntax. The reason to choose novels instead of other kinds of literary networks such as journals, dictionaries or other linguistic standards such as the British National Corpus \cite{25}, is because we believe that the process of writing a piece of art is in itself a trial to make the structure of the text more similar to the surrounding environment. It is probably the reason why well written novels display cleaner statistical behaviour. Moreover a novel can be considered as a ``closed" system in the sense that it contains all the information to understand it, for instance no references are needed. In this sense we believe a novel to be the best sample to consider in order to study the properties of language.

 The affinity of a system like a novel to systems such as food webs, or social systems, as we saw in this paper, can be surprising, but it has already been considered in philosophy. In fact, many philosophers consider social and linguistic systems to be characterized by a strict analogy.  In both systems, internal $elements$ or identities are linked by differential relations and can be $articulated$ - that is related - so as to become $moments$ of a specific articulated totality. This entails that the very process of articulation will modify their previous identity \cite{19}.

With this in mind, we algorithmically demonstrated that the scale free degree distribution and the scale free weight distribution of the analysed weighted networks are implied by the scale free strength distribution and that the reverse doesn't hold. We also noticed that in a network such as the linguistic one, that is not highly clustered,  the  behaviour of the average clustering coefficient and of the average nearest neighbour distribution doesn't change in the shuffling process. It's not the same in the case of the scientific collaboration network, where the clustering is an important growth fingerprint. In this case both the high average clustering coefficient and the assortative nearest neighbour degree behaviour are lost in the shuffling operation. This means that those properties are very important for the description of some specific networks, but not for all of them.

We then introduced the selectivity measure, defined as the average weight distributed on the links of a single vertex. We showed that this measure is able to statistically catch the local structures of  networks and so to easily distinguish a real network from a shuffled network. In the case of language large values for the selectivity indicate
tokens that have an exclusive relation with other tokens and that form morphological structures. In the case of the scientific collaboration network large values of the selectivity indicate authors that have exclusive relations with a few other authors. Via an ad-hoc stochastic network we showed that the selectivity of the vertices is very sensitive to the correlations in the system.

The fact that novels, investigated at the level of tokens, have long range fractal correlations  is already well known \cite{hl33}. Nevertheless vertex selectivity is not a measure of first order correlations.  The normalised weight of the links between tokens, or at least its deviation from the average weight,  should be  the best candidate for measuring such topological correlations. The problem is  that the scale-free distribution for the frequency of the tokens implies the invariance of the scale-free behaviour for the weight distribution so that it is very difficult to determine if the system is highly correlated or not (see left panel of Fig.\ref{f3}). In this sense vertex selectivity is not related to any of the classical correlation measures, it's just a measure that characterises the quality of the differential relations between pairs of elements in the system. Moreover it is worth  mentioning that the measure of the selectivity is not really clean. That is quite evident from the figures in which, even with a large logarithmic binning,  data don't align very well. In order to better understand the data and the finite size effects it would be interesting to carry out further research to examine the data using some of the more recent techniques for finite sample statistics \cite{hl34}.

We would like to stress that if the mean field approach is well defined for the calculation of the degree distribution in the Barabasi-Albert model, that is an unweighed tree network, then it is not applicable to  weighted networks without considering the  correlations naturally arising in the adjacency matrix. The usual approach tends to conflate the strength and the degree \cite{25} and the analytical results hold for the straight connection between the two measures.

The fact that  in the analysed networks  the degree distributions for real and shuffled networks are indistinguishable implies that in our empirical studies the degree distribution is implied by the strength distribution. The strength distribution represents the number of times an event appears or interacts in the system. This number has been considered in information theory and analysed via the Shannon entropy \cite{20}, the degenerate Shannon entropy \cite{21} and the Kolmogorov or $algorithmic$ entropy \cite{22}. The entropic approach to language requires one to maximise the amount of information of a system by minimising opportune quantities, usually via Lagrange multipliers, such as the receiver effort. It has been
demonstrated that this goal can be achieved if the distribution of the number of elements in the system is a power law \cite{23}. In particular, through  the degenerate entropy approach, that considers all the elements in the system with the same frequency as equivalent, it is possible to reproduce the exponential cutoff for small values of the strength in the strength distribution,  evident in Fig.\ref{f6} and in many other networks \cite{1}. Nevertheless, since the Shannon entropy is defined through the relative frequency of the elements of the system, this measure doesn't account of the  peculiar arrangements of the elements, that is if we shuffle the elements of the system, the resulting Shannon entropy or degenerate Shannon entropy doesn't change. A more sophisticated approach via information theory is suggested by the work of scientists dealing with food webs \cite{24}. In fact they define entropy based on the actual fluxes between trophic species, that is based on the weights of a weighted adjacency matrix. This type of approach takes into account the differential relations between the different elements of the system. Such an approach was recently considered in network theory in \cite{26}. Unfortunately, despite promising results, the analytical development of the approach is difficult.

\section*{Acknowledgments}
We thank the European Union Marie Curie Program NET-ACE (contract number MEST-CT-2004-006724) for financial support. We would also like to thank Andrea Mura for  many useful suggestions.


\thebibliography{apsrev}
\bibitem{18} S. Abe, N. Suzuki,  Small-world structure of earthquake network,  Physica A \textbf{337}, 357 (2004).
\bibitem{11} L. Antiqueira, M. das Gracas V. Nunes, O. N. Oliveira, L. da F. Costa, Strong correlations between text quality and complex networks features, Physica A \textbf{373}, 811 (2007).
\bibitem{1} A.L. Barabasi, R. Albert,  H. Jeong, Mean-field theory for scale-free random networks,  Physica A \textbf{272}, 173 (1999).
\bibitem{15} A.L. Barabasi, H. Jeong, Z. Neda, et al., Evolution of the social network of scientific collaborations,   Physica  A \textbf{311}, (2002).
\bibitem{8} A. Barrat, M. Barthelemy, R. Pastor-Satorras, et al., The architecture of complex weighted networks,  Proc. Natl. Acad. Sci. USA \textbf{101}, 3747 (2004).
\bibitem{3} I. Bose, B. Ghosh, R. Karmakar, Motifs in gene transcription regulatory networks,  Physica A \textbf{346}, 49 (2005).
\bibitem{17} C. Cattuto, C. Schmitz, A. Baldassarri, et al., Network properties of folksonomies,  AI Communications \textbf{20}, 245 (2007).
\bibitem{4} R.F. i Cancho, R. Sole', R. V. , The small world of human language, Proceed. of the Royal Society of London Series B \textbf{268}, 2261 (2001).
\bibitem{13} R.F. i Cancho,  R.V. Sole', Two regimes in the frequency of words and the origins of complex lexicons: Zipf's Law Revisited, Journ. of Quant. Ling. \textbf{8}, 165 (2001).
\bibitem{25} S.N. Dorogovtsev, J.F.F. Mendes,  Evolution of networks, Advances in Physics \textbf{51}, 1079 (2002).
\bibitem{bs} B. Efron, R.J. Tibshirani,  \emph{An Introduction to the Bootstrap}, Chapman and Hall, New York (1993).
\bibitem{hl32} J. Kerouac, \emph{On the road}, Penguin Classics (2002).
\bibitem{22} A.N. Kolmogorov, Combinatorial foundations of information-theory and the calculus of probabilities,  Russian Math. Surveys \textbf{38}, 29 (1983).
\bibitem{19} E. Laclau, C. Mouffe, \textit{Hegemony and Socialist Strategy}, (London and New York, Verso 1985).
\bibitem{23} B. Mandelbrot, Information theory and psycholinguistics: a theory of word frequencies, Reading in Math. Soc. Sciences, The M.I.T. Press, 350 (1966).
\bibitem{12} A.P. Masucci, G.J. Rodgers, Network properties of written human language, Phys.Rev.E \textbf{74}, 026102 (2006).
\bibitem{6} A.P. Masucci, G.J. Rodgers,	Multi-directed Eulerian growing networks, Physica A \textbf{386},  557 (2007).
\bibitem{2} A.P. Masucci, G.J. Rodgers, The network of commuters in London, accepted for publication by Physica A, preprint at http://arxiv.org/abs/0712.1960 (2007).
\bibitem{5} H. Melville,  \textit{Moby Dick: or, the whale}, (Penguin Popular Classics, 1994).
\bibitem{hl33} M.A. Montemurro, P.A. Pury, Long-range fractal correlations in literary corpora, Fractals \textbf{10}, 451 (2002).
\bibitem{21} S. Naranan, V.K. Balasubrahmanyan, Information theoretic models in statistical linguistics 1. A model for word frequencies, Current Science \textbf{63}, 261 (1992). 	
\bibitem{16} M.E.J. Newman,  The structure of scientific collaboration networks, Proceed. of the Nat. Acad. of Scien. of the Un. St. of Am. \textbf{98}, 404 (2001).
\bibitem{pl} M.E.J Newman, Power laws, Pareto distributions and Zipfs law, Contemporary Physics \textbf{46}, 323 (2005).
\bibitem{hl6} G. Orwell, \emph{Nineteen Eighty-four}, Penguin Books Ltd Paperback, (1990).
\bibitem{hl34} T. P$\ddot o$schel, W. Ebeling, C. Fr$\ddot{o}$mmel, R. Ram\'{y}rez, Correction algorithm for finite sample statistics, Eur. Phys. J. E \textbf{12}, 531 (2003).
\bibitem{20} C. Shannon, W. Weaver, \textit{A mathematical theory of communication}, (University of Illinois Press, Urbana, 1949).
\bibitem{14} A. Spencer, \textit{Morphological theory: an introduction to word structure in generative grammar}, (Oxford: Blackwell, 1991).
\bibitem{24} R.E. Ulanowicz, W.F. Wolff, Ecosystem flow networks - loaded dice,  Math. Biosci. \textbf{103}, 45 (1991).
\bibitem{26}T. Wilhelm, J. Hollunder,  Information theoretic description of networks, Physica A \textbf{385}, 385 (2007).
\bibitem{9} G.K. Zipf, \textit{Human behaviour and the principle of least effort}, (Addison-Wesley Press, 1949).
\bibitem{10} A collection of papers dedicated to network theory applied to human language can be found at http://www.lsi.upc.edu/~rferrericancho/linguistic\_and\_cognitive\_networks.html .
\bibitem{7} \textit{Web of Science}, http://portal.isiknowledge.com/ .
    \end{document}